\documentclass[aip,rsi,groupedaddress,reprint,graphicx]{revtex4-1}
\pdfoutput=1
\usepackage{amsmath}
\usepackage{amssymb}
\usepackage{graphicx}
\usepackage{dcolumn}
\usepackage{bm}

\begin{document}

\title{Rugged constant-temperature thermal anemometer}

\author{J. Palma}
\altaffiliation{Present address: Comercial Evolux Ltda. Av. Padre Hurtado Central 1298, Las Condes, Santiago, Chile.}
\affiliation{Laboratorio de Turbulencia, Departamento de F\'isica,
Facultad de Ciencia, Universidad de Santiago de Chile, Casilla
307, Correo 2, Santiago, Chile}

\author{R. Labb\'e}
\email[]{raul.labbe@gmail.com}
\affiliation{Laboratorio de Turbulencia, Departamento de F\'isica,
Facultad de Ciencia, Universidad de Santiago de Chile, Casilla
307, Correo 2, Santiago, Chile}

\date{\today}

\begin{abstract}
Here we report a robust thermal anemometer which can be easily built. It was conceived to measure outdoor wind speeds, and for airspeed monitoring in wind tunnels and other indoor uses. It works at a constant, low temperature of approximately $90~^\circ$C, so that an independent measurement of the air temperature is required to give a correct speed reading. Despite the size and high thermal inertia of the probe, the test results show that this anemometer is capable of measuring turbulent fluctuations up to $\sim 100$~Hz in winds of $\sim 14$~m/s, which corresponds to a scale similar to the length of the probe.\\\\ \copyright {\it 2016 American Institute of Physics. This article may be downloaded for personal use only. Any other use requires prior permission of the authors and the American Institute of Physics. This article appeared in} Rev. Sci. Instr. {\bf 87}, 125112 (2016) {\it and may be found at} http://dx.doi.org/10.1063/1.4972585
\end{abstract}


\maketitle

\section{Introduction}
\label{Intro}
For many decades, thermal anemometers have been employed in a diversity of applications where the speed of a fluid (or its gradient) needs to be measured. In the field of turbulent flows the hot wire anemometer (HWA), and particularly its constant temperature version (CTA), have been the preferred choice due to its excellent spatial resolution and, in the case of the latter, its unchallenged frequency response. Some of the disadvantages of the HWA are its fragility ---the wire diameter is typically about $5~\mu$m--- and its susceptibility to surface contamination, which makes necessary frequent calibrations to maintain its accuracy. For industrial applications we find a variant of the HWA in which the heating element is covered with a protective ceramic or other appropriate material, to avoid damage of the heating element by chemical compounds or solid particles. The latter type of device is adequate for applications demanding a robust probe, like outdoor measuring of wind speed, or airspeed measuring in wind tunnels or other devices for monitoring/control purposes. Due to the protective covering, the thermal mass of these probes is bigger, and the thermal conductivity of the heating element to the surroundings is smaller. Therefore, a wide frequency response cannot be expected from this type of probe. Nevertheless, in the constant temperature mode a bandwidth up to $100$~Hz is achievable, as we will find out later. Other types of probes for thermal anemometry will be briefly mentioned in the Appendix \ref{Ap_A}.

The remaining parts of this article are organized as follows: in Section \ref{Theory} we will depict the theoretical aspects required for the design of a working CTA, starting with the justification for using a proportional-integral (PI) controller, and give the results of some simulations. In Section \ref{EC} the schematic diagram of the electronic circuit of the servo-controller, which is nothing more than the physical realization of the dynamical equations of the Section \ref{Theory}, will be presented. In Section \ref{M_E} details for the construction of the probe are given, while the results of the tests performed on the constructed device are given is Section \ref{Test}. Lastly, in Section \ref{Concl} the conclusions are given.

\section{Theory and numerical simulations}
\label{Theory}

The theory of thermal anemometers is well known, so that we will adopt a mostly practical approach, paying attention mainly to aspects related to our specific goal. Initially, and for a long time, commercial and laboratory CTAs used proportional controllers to keep a constant sensor temperature. Today, some models using a PI controller can be acquired in the market. This type of controller can also be found in new developments, like the constant bandwidth CTA reported by Ligeza.\cite{lig07,Lig09} The design described here also makes use of a PI controller. At first glance, the utilization of a PI controller seems to be a complication, because a proportional controller appears as being the simplest option to implement a servomechanism. In fact, that is not always the case. The reason is that a proportional controller needs a huge gain in order to attain a small error. Even if a thermal anemometer with proportional controller is carefully calibrated prior to use, its inherently imperfect bridge balance will lead to errors in the working temperature and, consequently, in the speed measurement. Thus, a high gain is mandatory to obtain a small error, and imposes the utilization of fast electronics. This aspect was studied by Freymuth,\cite{Frey67,Frey77,Frey77a} who in addition showed that the system of differential equations governing the whole system is of third order due to the two main poles usually found in voltage amplifiers. Indeed, the order can be even higher, depending on the number of cascaded stages used in the amplifier circuit. This is due to the upper bound in the gain-bandwidth product inherent to every voltage-gain stage of any amplifier. Thus, the higher the number of stages in the amplifier, the higher the number of poles and, consequently, the order of the governing equations. A third (or higher) order system with high gain, working in closed loop, will almost certainly be prone to instability, requiring compensating networks to avoid instabilities and, hopefully, obtain a critically damped system. In his development of an algorithm for deriving the transfer functions of hot-wire CTAs of arbitrary complexity, Watmuff\cite{Watmuff95} states that insufficient amplifier bandwidth is one of the primary sources of instability in hot wire anemometers. In the discussion that follows it is implicitly assumed that, for small perturbations around the equilibria, the anemometer can be seen as a linear system. It is in that context that the terminology of linear control systems is used.

In the present application, the controller goal is to maintain a perfectly balanced Wheatstone bridge, at least within a frequency band useful for wind speed measurements and wind tunnel control. Due to its operating principle, a first order controller that can perform that task is the PI controller. There are two reasons that justify this choice. On the one hand, the integral stage allows for a perfect balance of the bridge for static, or quasi-static inputs. On the other hand, the proportional stage compensates the unbalance coming from varying inputs with moderate bandwidth, without the need of a huge gain, because the quasi-static part of the error is already compensated by the integral stage. Given that moderate gains for the amplifiers are enough in this configuration, the contribution of their poles to the controller transfer function is negligible. Thus, the only relevant derivatives in the governing equations are those resulting from the integrator and the thermal response of the probe, which have also a first order dynamics. In other words, we need to deal only with a second order system, which is simpler than the third (or higher) order system resulting from a high gain proportional controller. We must stress that this somewhat paradoxical result is due simply to the fact that the bandwidth limitations in the amplifier of a proportional controller come into play owing to the need of a high gain.
\begin{figure}[t]
\centering \vspace{0.2cm} \hspace{-0.2 cm}
\includegraphics[width=.46\textwidth]{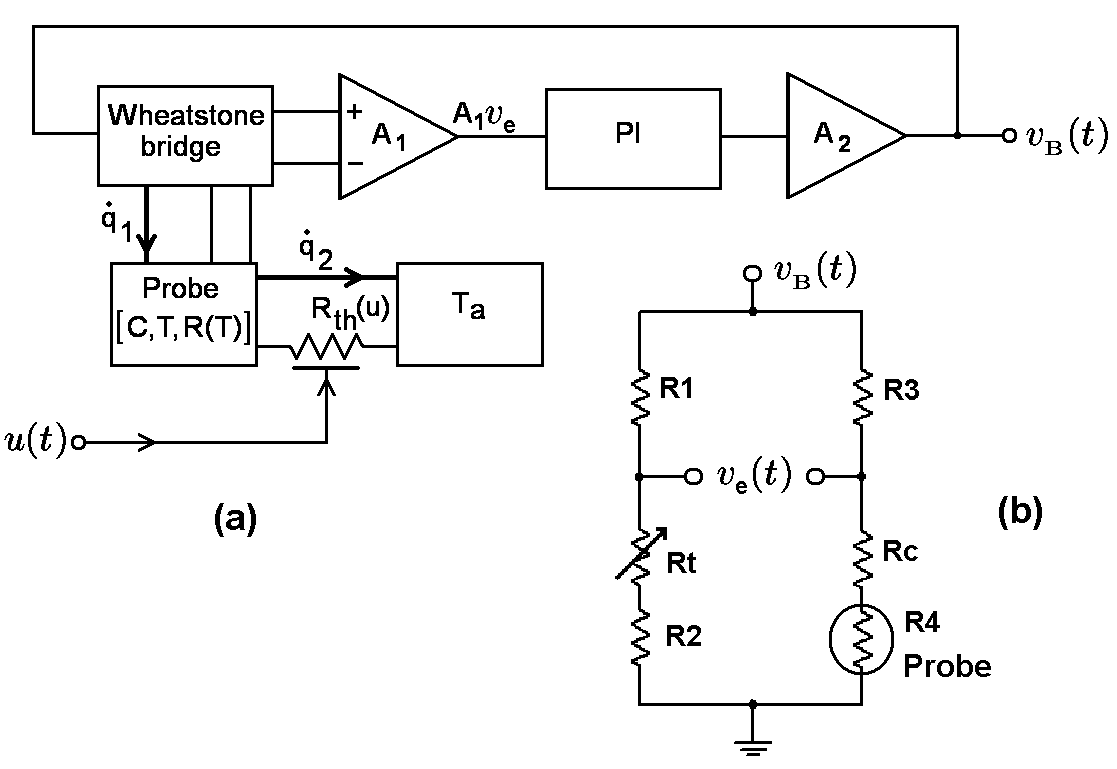}
\caption{(a) Block diagram of the thermal anemometer. The input signal is the airspeed $u(t)$, while the voltage applied to the bridge,  $v_\mathrm{_B}(t)$, is the output. The rate of heat flow $\dot{q_1}$ is provided by the Joule dissipation at the probe resistor, while $\dot{q_2}$ represents the rate of heat flow towards the surroundings at temperature $T_\mathrm{a}$, through the thermal resistance $R_\mathrm{th}(u)$, whose value depends on $u(t)$. (b) Schematic diagram of the Wheatstone bridge, showing the relationship between components and voltages (see text).}\label{one}
\end{figure}

Figure \ref{one}(a) displays a block diagram of the thermal anemometer. There, $u(t)$ is the wind speed, and plays the role of system input, whereas $v_\mathrm{_B}(t)$ is the voltage applied to the bridge, and also the system output. Figure \ref{one}(b) shows the details of the Wheatstone bridge, where $R_4$ is the probe. For maximum sensitivity, one should have $R_1=R_2$ and $R_3=R_4$. The resistor $R_c$ represents the resistance of the cable that connects the probe to the bridge. Its resistance can degrade the anemometer performance to an unacceptable level, so that it must be minimized as much as possible. The variable resistor $R_t$ compensates the cable resistance, restoring the operating temperature of the probe to the desired value. Its resistance must be set to
\begin{equation}
R_t = \frac{R_2 Rc}{R_3},
\label{Rcomp}
\end{equation}
where $R_3$ is equal to the probe resistance at the operating temperature.

Note that the input $u(t)$ modulates the thermal resistance $R_{th}(u)$, which is a system parameter that controls the rate of heat flow $\dot{q}_2$ leaving the probe towards the thermal reservoir $T_a$. Thus, in thermal anemometers the input is parametrically coupled to the system. Next, the rate of heat flow $\dot{q}_1$ entering the probe is provided by the Wheatstone bridge as a fraction of the power generated by $v_\mathrm{_B}(t)$. Again, this interaction is parametric, and quadratic in $v_\mathrm{_B}(t)$. Thus, this system will be governed by a system of two non linear, first order equations, with parametric coupling. We will solve it by using numerical integration.

Let us start with the well known equation for the probe thermal balance: the electrical power entering the probe, and the rate of heat flow leaving the probe towards the surrounding air, determine the rate of change in the probe temperature. If $C$ is the probe heat capacity and $T$ its temperature, then $C\dot{T}= \dot{q}_1 + \dot{q}_2$. Assuming that the probe resistance changes linearly with the temperature, we have
\begin{equation}
R_4=\alpha T,
\label{R_T}
\end{equation}
with $\alpha$ constant. This probe is connected to the controller through a cable, which must be chosen as short as possible. As this anemometer is designed to measure mainly the mean value of the airspeed, or at most fluctuations in a rather low frequency band, the cable inductance and capacitance are not of concern. However, the cable resistance could be a problem, given the low value of the probe resistance. Thus, in our equations we will consider the cable resistance, $R_c$. Now, let $Y(u)\equiv R^{-1}_{th}(u)$ be the thermal conductance to the surroundings when the wind speed is $u$. Then, given that $v_\mathrm{_B}$ is the voltage applied to the bridge, we have
\begin{equation}
C\dot{T} = \frac{\alpha T v^2_\mathrm{_B}}{(R_3 + \alpha T + R_c)^2} + Y(u)(T_a - T), \label{Probe}
\end{equation}
where $R_3$ is the bridge resistor in series with the probe, $T$ is the probe temperature, and $T_a$ is the air temperature at a distance much larger than the size of the probe. Let $v_e(t)$ be the bridge imbalance voltage, or error voltage. The PI controller output, which is also the voltage applied to the bridge, is given by
\begin{equation}
v_\mathrm{_B}(t) = P v_e(t) + I\int_{t_0}^t v_e(\xi)d\xi, \label{PI}
\end{equation}
with
\begin{equation}
v_e(t) = \biggl(g - \frac{\alpha T + R_c}{R_3 + \alpha T + R_c}\biggr)v_\mathrm{_B}(t), \label{Verr}
\end{equation}
where $g=(R_2+R_t)/(R_1+R_2+R_t)$, and $R_1$, $R_2$ are the resistances of the resistors in the reference arm of the bridge, as shown in figure \ref{one}(b). The parameters $P$ and $I$ are the proportional and integral gains, respectively, and include the gains of the amplifiers A$_1$ and A$_2$. Equations (\ref{Probe}) through (\ref{Verr}) govern the system dynamics, and the anemometer response can be optimized by adjusting the parameters $P$ and $I$. Replacing
$v_e(t)$ in eq. (\ref{PI}) and taking the derivative, we obtain a first order equation that involves only the voltage $v_\mathrm{_B}(t)$. After some straightforward algebra, and defining
\begin{equation}
h = \frac{1}{C}\biggl[\frac{\alpha T v^2_\mathrm{_B}}{(R + \alpha T)^2} + Y(u)(T_a - T)\biggr], \label{Teq}
\end{equation}
with $R=R_3+R_c$, we obtain the dynamical equations for the thermal anemometer with a PI controller:
\begin{subequations}
\label{DynEq}
\begin{equation}
\dot{T} = h, \label{DynEq:Temp}
\end{equation}
\begin{equation}
\dot{v}_\mathrm{_B}=\frac{\displaystyle I\biggl(g-\frac{R_c + \alpha T}{R+\alpha T}\biggr)-P\frac{\alpha R_3h}
{(R+\alpha T)^2}}{1-\displaystyle P\biggl(g-\frac{R_c + \alpha T}{R+\alpha T}\biggl)}~v_\mathrm{_B}.\label{DynEq:Vout}
\end{equation}
\end{subequations}
Note that the exact form of the thermal conductance $Y(u)$ is not relevant: if $Y_0$  and $Y_{\mathrm{max}}$ are the thermal conductances at zero and maximum speeds,  respectively, small departures of $Y(u)$ from the ideal thermal conductance function, while keeping the mapping of the input speed range on the interval $[Y_0,Y_{\mathrm{max}}]$, will give essentially the same results. Given that this system is nonlinear, optimization cannot be achieved on the whole operating range unless an adaptive controller is used. Our approach was to optimize the anemometer for some speed $u$, $0 < u < u_{\mathrm{max}}$, to obtain an acceptable behavior within the entire operating range.
\begin{figure}[t]
\centering \vspace{0.2cm} \hspace{-0.2 cm}
\includegraphics[width=.48\textwidth]{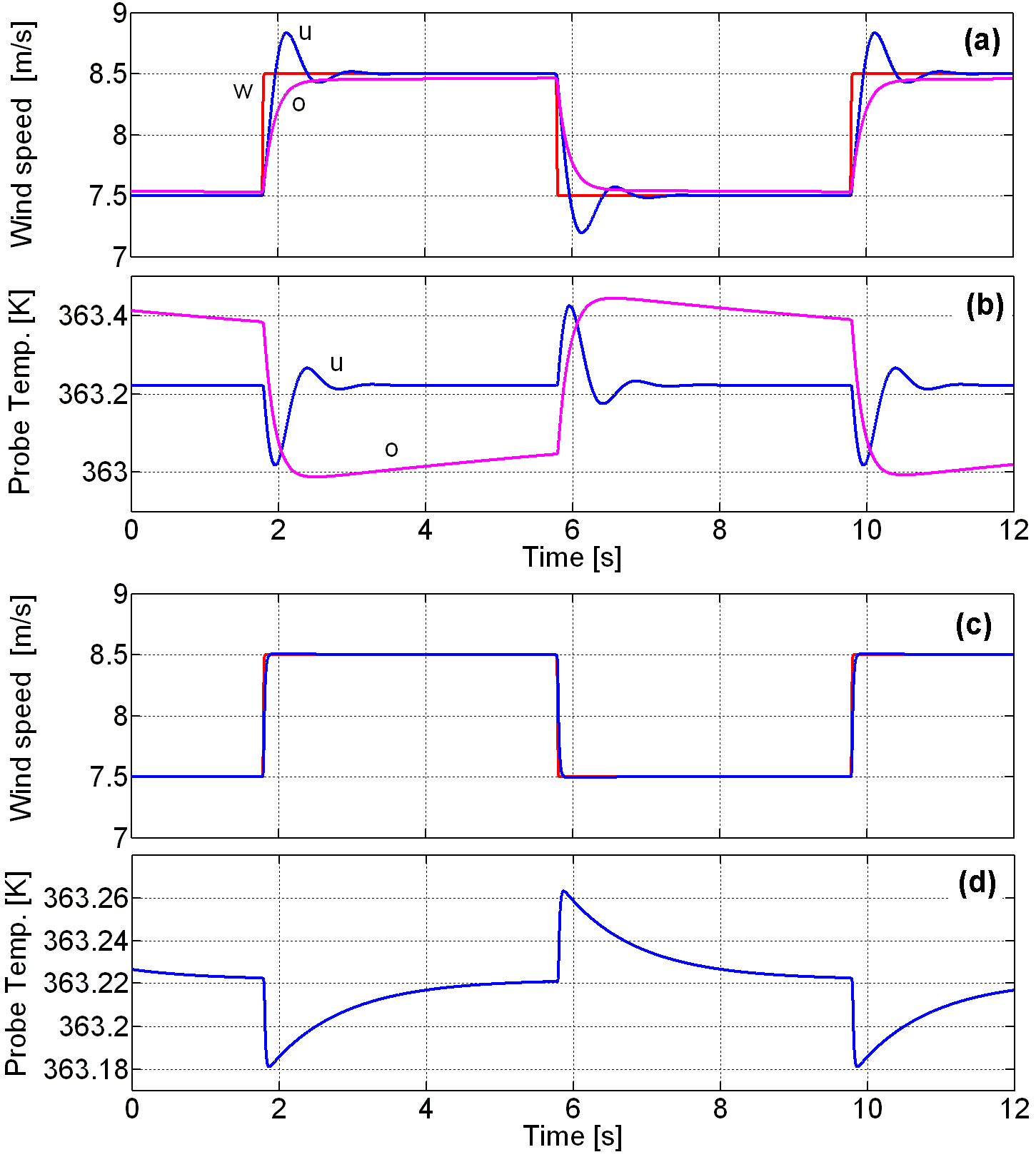}
\caption{(Color online). Anemometer theoretical responses to a wind (w, red) with average speed $v_{_A}=8.0$~m/s and a square fluctuation of amplitude $v_{_F}=0.5$~m/s for two settings of the parameters $P$ and $I$. (a) The measured wind speed displays an underdamped response (u, blue) when $P=100$ and $I=1000$, whereas an overdamped response (o, magenta) is obtained when $P=100$ and $I=10$. In (b), a plot of the corresponding fluctuations in the probe temperature is displayed. In the optimal case, $T_{probe}$ should be constant. (c) For $P=1000$ and $I=1000$ the response seems to be nearly optimal. The fluctuation amplitude of $T_{probe}$ (d) is a fifth of the amplitude of the curves displayed in (b).} \label{FWind}
\end{figure}

To solve the equations (\ref{DynEq}), the Bulirsh-Stoer method\cite{PreTeuVet97} can be used. To obtain a reference behavior, let us consider the ideal case with $R_c=R_t=0$. The input signal will represent a wind with average speed $u_0 = 8.0$~m/s, and a superimposed square wave of amplitude $u_\mathrm{m} = 0.5$~m/s. To control the steepness of the transitions between the two wind levels, the square wave can be approximated using a smooth function,
\begin{equation}
u_\mathrm{sw}(t) = u_\mathrm{m} \tanh\bigl[\beta \sin(2\pi t/\tau)\bigr],
\label{Wind}
\end{equation}
where $\tau$ is the period, and $\beta$ is the steepness parameter. Thus, the system input is given by $u(t)=u_0+u_\mathrm{sw}(t)$. Three solutions of the equations (\ref{DynEq}), obtained using three parameter settings, are displayed in the Figure \ref{FWind}. The curves in subplot \ref{FWind}(a) represent the speed theoretically measured by the anemometer for a wind having mean speed of $8.0$~m/s, with a symmetric square wave fluctuation of amplitude $0.5$~m/s. The input wind speed is represented by the square signal (w, red). The curve showing under damped response (u, blue) results with a parameter setting $P=100$ and $I=1000$, while the overdamped response (o, magenta) is obtained when $P=100$ and $I=10$. The corresponding thermal responses of the probe are displayed in the subplot \ref{FWind}(b). We see that in both cases the probe temperature deviates from the working temperature by about $0.2$~K. When $P=1000$ and $I=1000$, a nearly critical damping is obtained, as displayed in subplot \ref{FWind}(c), where the input and measured winds are almost superposed. In this case, as shown in subplot \ref{FWind}(d), the probe temperature deviates from the working value by only about  $0.04$~K, an amount five times smaller than those in the two previous cases. In principle, by increasing even more both, $P$ and $I$, an even better response should be obtained. The problem is that our model does not take into account the limitations of the operational amplifiers. Thus, what could appear as a nearly perfect performance in the model---when the parameter are set at very large values---can result in practice in a potentially destructive oscillation.

If the parameter values used to obtain the critically damped response are held fixed, and the mean speed of the wind is changed, a small change in the response is observed. With a mean wind speed of $15$~m/s the overall system response improves. The measured speed follows more closely the square wave, and the deviations of the probe temperature are smaller. Contrarily, when the mean wind speed is reduced to only $1.5$~m/s, an overshot of about $4$\% appears in the measured speed, and the deviation of the probe temperature is about $0.23$~K, that is, around $5.7$ times larger than that observed when the mean speed is $8.0$~m/s. This behavior is characteristic of the CTA, and is due mainly to the dependence of the probe-ambient thermal conductivity on the wind speed.

In the previous simulations the effect of the cable resistance, $R_c$, was neglected. This is not a problem when $R_c$ is much smaller than the resistance of the sensor. However, for $R_c$ values about 20\% the probe resistance, the degradation of the anemometer performance can be unacceptable. This problem is effectively solved by the variable resistor $R_t$ in series with $R_2$. Simulations using $R_c=0.5$~$\Omega$ and a value of $R_t$ calculated using the equation (\ref{Rcomp}) give results practically identical to those obtained with $R_c=0$. In fact, the results cannot be exactly the same: although the compensating resistor $R_t$ restores the operating temperature of the probe, the presence of $R_c$ and $R_t$ degrade the bridge sensitivity. This reduces the loop gain, and the signal to noise ratio.

\section{Electronic circuit}
\label{EC}

\begin{figure}[t]
\centering \vspace{0.2cm} \hspace{-0.2 cm}
\includegraphics[width=.48\textwidth]{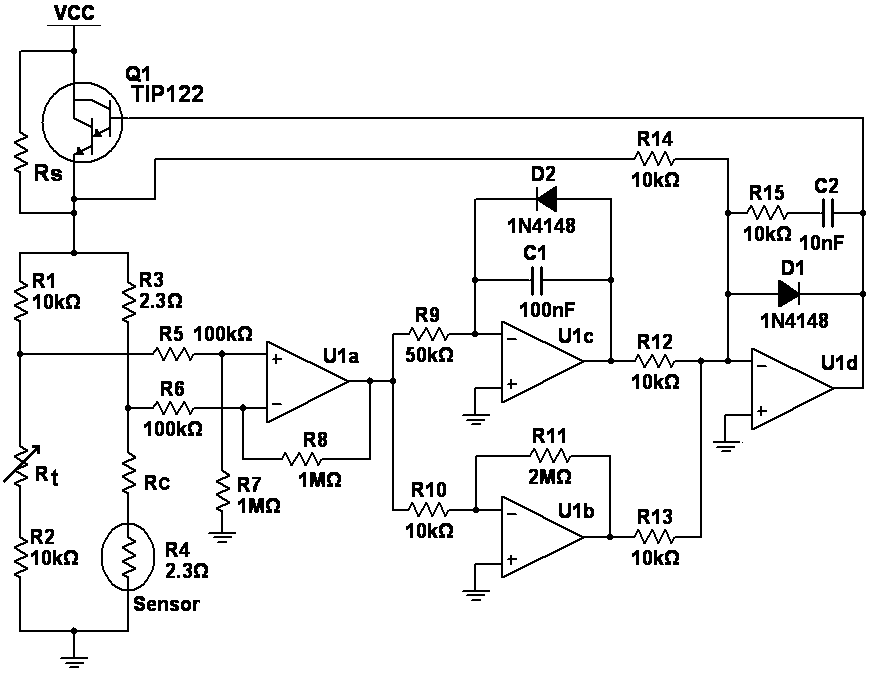}
\caption{{Schematic diagram of the anemometer electronics (resistances in Ohm). A three-stage PI controller provides the right amount of power to keep the probe temperature at the prescribed value. The error voltage is amplified by the first stage (U1a), and amplified and integrated in the second stage (U1b, U1c). Next, the proportional and integral components are added and current boosted by the third stage (U1d, Q1), whose output powers the bridge (see text). The variable resistor $R_t$ in its reference arm compensates the resistance $R_\mathrm{c}$ of the cable connecting the probe to the bridge.}} \label{Elec}
\end{figure}

A schematic diagram of the electronic circuit is displayed in the Figure \ref{Elec}. The Wheatstone bridge consists of four resistors, $R_1$ through $R_4$, where $R_4$ is the sensor. The resistances of the reference arm are $R_1=R_2=10$~k$\Omega$, while on the probe arm the upper resistance is $R_3=2.3$~$\Omega$, which is the resistance value to be attained by the sensor at the working temperature (or equilibrium point). Thus, at the equilibrium point the voltage at the nodes in each arm will be half the voltage applied to the upper node of the bridge. The controller has three stages: the first one is a difference amplifier, U1a, which amplifies the bridge output. This voltage ---the error voltage--- is multiplied by ten and delivered to the second stage, or PI stage, where it is amplified by U1b and simultaneously integrated by U1c. These voltages are added in the third stage, U1d, and current boosted by a Darlington transistor, Q1. This stage is simply a unity gain voltage amplifier with increased current source---but not sink---capability. Finally, the emitter of Q1 is connected to the upper node of the bridge, thus closing the loop. Diode D2 limits the negative excursion of the output of U1d to $\sim -0.7$~V. This prevents the application of large negative voltages to the base of Q1 during the power-on transient. The diode D1 has also a limiting function, preventing large positive voltages at the output of the integrator. If such condition is allowed, then the anemometer could remain latched in a state where the proportional stage in unable to deliver enough voltage to counteract the integrator output, and the bridge would never become energized. A resistor can optionally be placed in parallel with the pass transistor Q1. In the schematic diagram it is designated as $R_\mathrm{s}$. This resistor is useful when a small offset voltage makes the output of U1a negative, which locks the circuit in a state in which the bridge voltage remains null. Under such condition, the resistor $R_\mathrm{s}$ must provide enough current through the bridge for trespassing the $v_\mathrm{_B}=0$ equilibrium point towards the unstable branch. From here, the bridge voltage should transit to the next stable point, where the voltage at the nodes of both arms is half the voltage applied to the upper node of the bridge. Note that when a proportional controller is used, the resistor $R_\mathrm{s}$ introduces a systematic error in the CTA output. In the case of the PI controller, this error is automatically compensated by the integrator.
\begin{figure}[t]
\centering \vspace{0.2cm} \hspace{-0.2 cm}
\includegraphics[width=.40\textwidth]{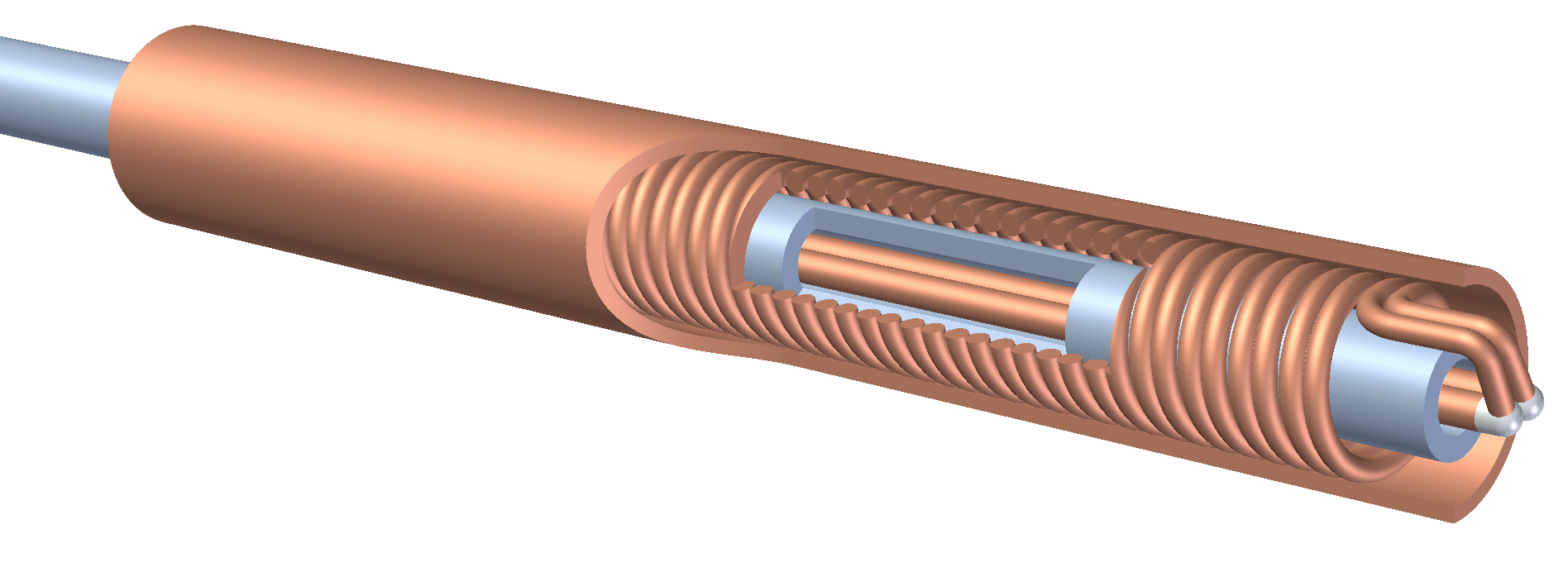}
\caption{{(Color online) Drawing of the anemometer probe. Some cuts were made to allow a view of the inner structure. A stainless steel tube supports a non inductive winding of isolated copper wire. An external copper tube acts as housing and temperature homogenizer. The assembly is filled with high temperature silicone glue (see text).
}} \label{Anem}
\end{figure}

\section{Probe details}
\label{M_E}

This CTA is based on a heater made of a non inductive winding of AWG~$36$ magnet wire. The idea is to take advantage of its electrically isolating film to protect the conducting element. This requires a low working temperature to avoid burning the coating. Thus, we chose a temperature close to $90^\circ$C. This temperature is well below the maximum allowed temperature for common enamel types, although for long durability at least a thermal class $180^\circ$C NEMA MW 30-C insulation should be used. To minimize the inductance of the probe, the winding was double-strand wound by hand around a hypodermic needle of diameter $1.0$~mm, to obtain a winding length of $\approx 35$~mm. A hollow cylinder made of $0.2$~mm thick copper sheet was used to enclose the winding and homogenize the temperature to avoid hot spots. High temperature silicone was used to fill the inner volume and improve the thermal contact between the wire and the enclosure. Figure \ref{Anem} displays the arrangement of metal components making the probe. Several cuts were made to the different parts to show the inner structure. The resistance of the finished probe at ambient temperature was $R_4\approx 1.8 \Omega$, whereas at the working temperature of $90^\circ$C $R_4\approx 2.3 \Omega$. In the finished probe the winding is completely covered by the copper tube.\\

\section{Test results}
\label{Test}

\begin{figure}[t]
\centering \vspace{0.2cm} \hspace{-0.2 cm}
\includegraphics[width=.47\textwidth]{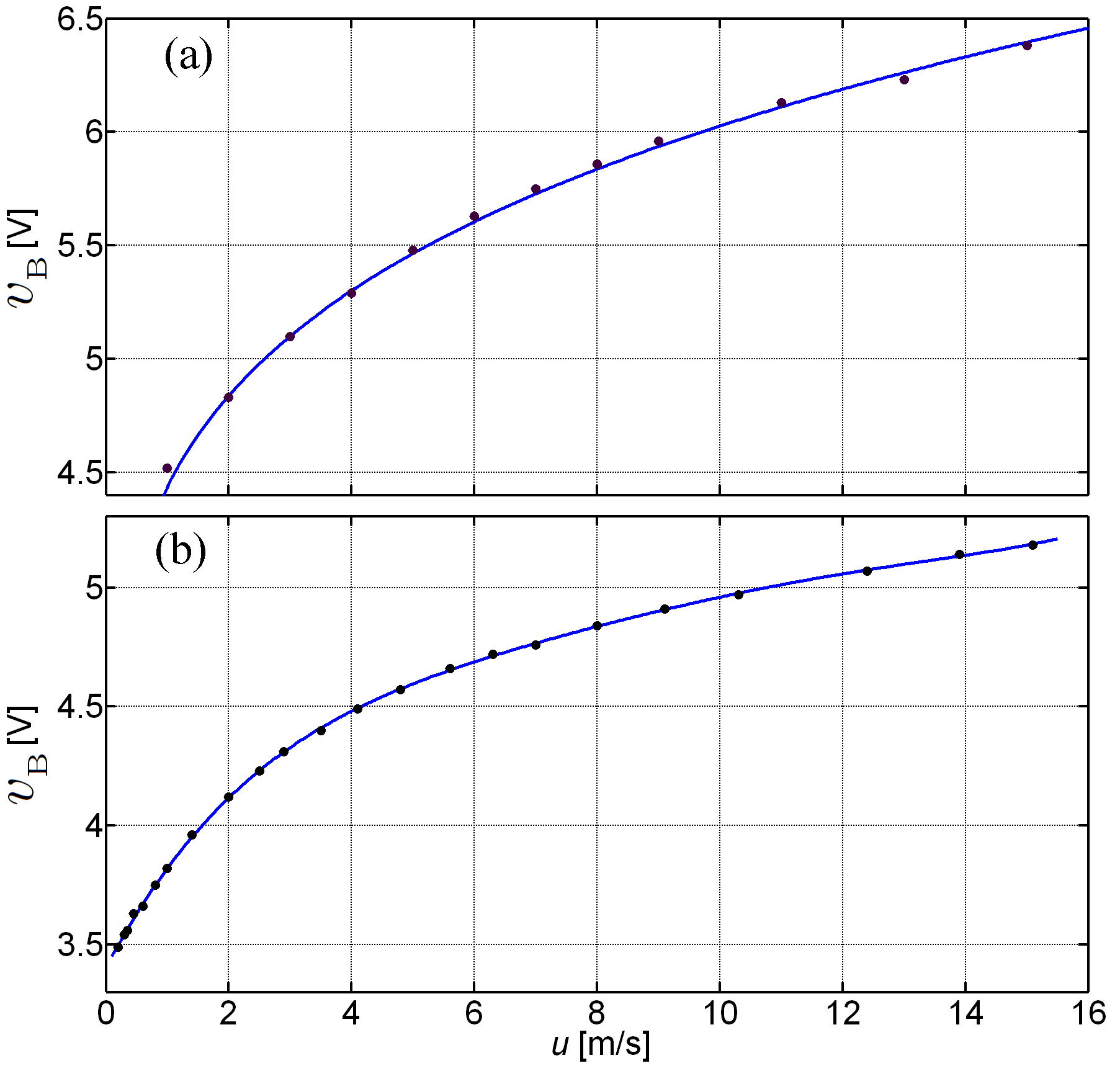}
\caption{{(Color online) Two examples of calibration. (a) Fit of the bridge voltage to the jet airspeed data using the King's law. The continuous curve represents the voltage given by equation (\ref{KL}), while the filled circles correspond to the measured data. (b) In this case, the gain of the controller and the cable resistance were reduced to improve the system stability. More points were used in the low speed range, starting at $u = 0.2$~m/s. Here, the continuous curve represents a fit using a fifth degree polynomial.}} \label{KLP}
\end{figure}
\begin{figure}[t]
\centering \vspace{0.2cm} \hspace{-0.2 cm}
\includegraphics[width=.46\textwidth]{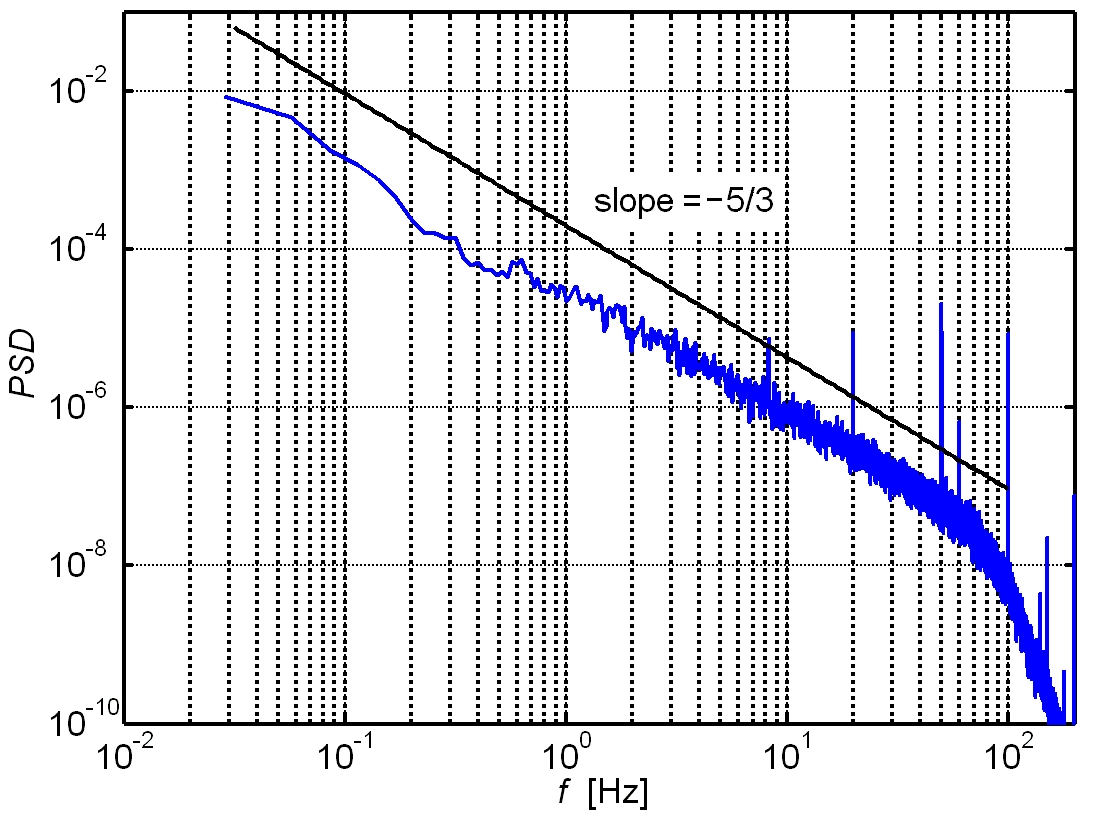}
\caption{{(Color online) Spectrum of the measured airspeed. At low
frequencies, the probe captures the turbulent component of the jet.
}} \label{Spectrum}
\end{figure}

The anemometer was tested using a square jet with section $14$~cm~$\times 14$~cm, with a turbulence level of about $5$\%. The parameter values were those of the components in the Figure \ref{Elec}, giving loop gain values $P=2000$ and $I=2000$. A rotating vane anemometer was used to measure the airspeed of the jet. The measurements were made within a speed interval from $1.0$~m/s through $15.0$~m/s. The plot in Figure \ref{KLP}(a) displays the speed measured with the vane anemometer, along with the calibration curve obtained by fitting the King's law to the bridge voltage. If $v_\mathrm{_B}$ denotes the bridge voltage and $u$ is the jet airspeed, then
\begin{equation}
v_\mathrm{_B} = \sqrt{A + B u^q}, \label{KL}
\end{equation}
where $A$, $B$, and $q$ are the fitting parameters.
The expression in equation (\ref{KL}) can be inverted to obtain the airspeed $u$ as a function of the bridge voltage $v_\mathrm{_B}$. In fact, when the measurements are to be done in a restricted speed range, the calibration accuracy can be improved by fitting the inverse function
\begin{equation}
u = \biggl(\frac{v_\mathrm{_B}^2-A}{B}\biggr)^{\frac{1}{q}}.
\label{IKL}
\end{equation}
The plot in Figure \ref{KLP}(b) represents a calibration using reduced loop gains. The cable resistance was also reduced. In this case we expect a reduction in the bandwidth along with an increase in the loop stability. In this case, a fifth degree polynomial was used to fit the data, giving less error than the King's law.
Figure \ref{Spectrum} displays the spectrum of the velocity signal measured by the anemometer. As can be seen, the curve approximately follows a straight line having a slope $-5/3$, which corresponds to the Kolmogorov spectrum of a turbulent velocity field. At first sight, this finding could seem rather surprising, given the characteristic sizes of the probe. In fact, the measurement was made with the main component of the velocity perpendicular to the longitudinal axis of the probe, whose diameter and length are $D \approx 1.7$~mm and $l \approx 35$~mm, respectively. By using the Taylor hypothesis and the biggest probe dimension, we can estimate the cutoff frequency of this probe in a wind of $\sim 8$~m/s, impinging orthogonally on the probe. The results is $f_\mathrm{c}\approx 115$~Hz. Of course, this is roughly the maximum geometric cutoff. As the thermal inertia also plays a role, what we see in reality is a frequency response limited by the signal to noise ratio. The roll-off near $100$~Hz in the Figure \ref{Spectrum} is related to the filter used to suppress the high frequency noise. Without this filter, what we have is a spectrum that becomes nearly flat beyond $100$~Hz, contributing with nothing but noise to the anemometer output.

In the previous test, the parameter values were those defined by the values of the components in the Figure \ref{Elec}, giving loop gain values $P=2000$ and $I=2000$. This setting worked like the simulation in the first anemometer we built. However, in the second one an oscillation was observed. This was corrected by reducing the value of the integral gain.

\begin{figure}[t]
\centering \vspace{0.2cm} \hspace{-0.2 cm}
\includegraphics[width=.47\textwidth]{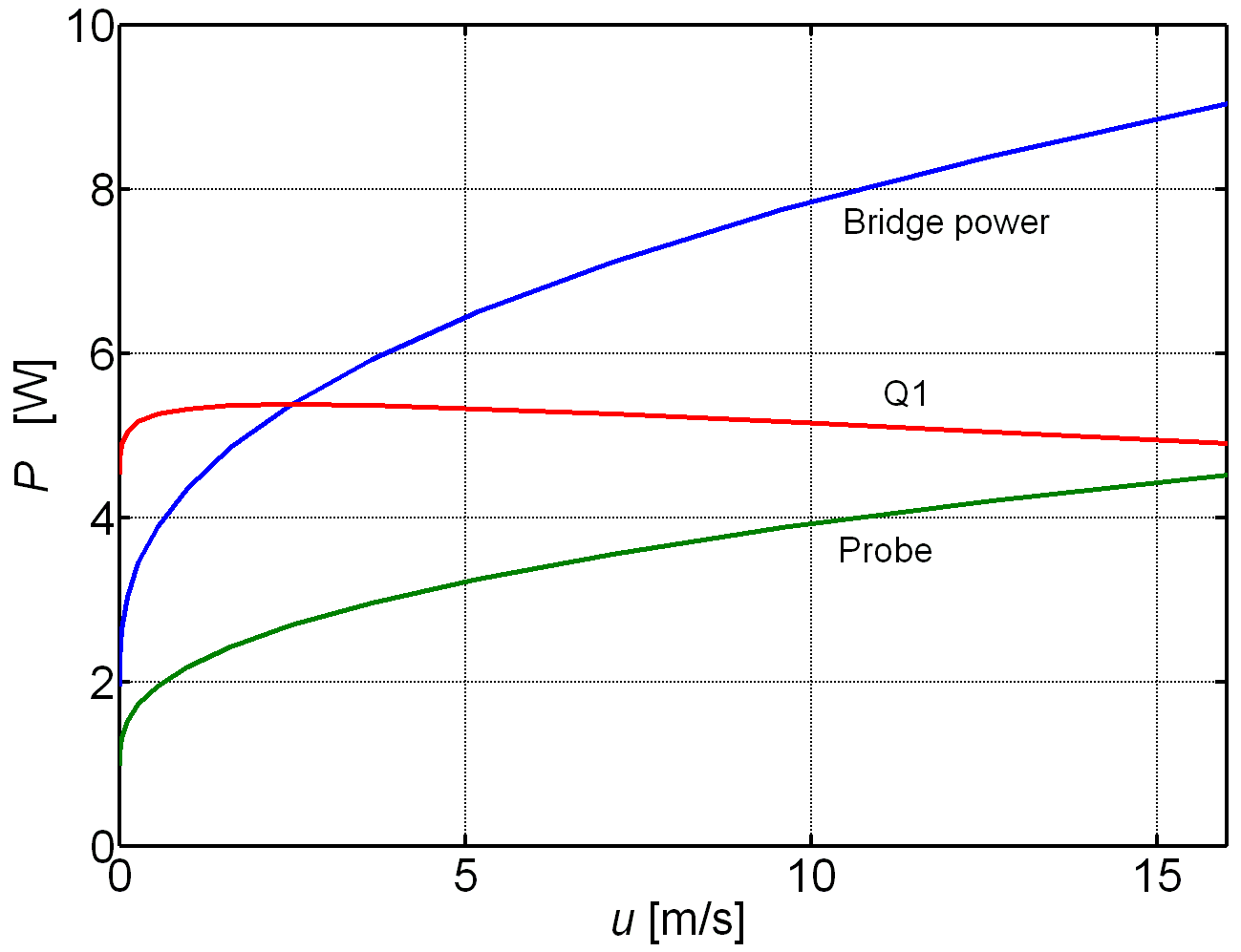}
\caption{{(Color online) Electrical power dissipated by the main components of the bridge when the power supply voltage is $V_\mathrm{CC}=10$~V. At $u=16$~m/s the total power consumption is $9$~W. The dissipation in the pass transistor, $Q_1$, is $P_\mathrm{_T}\gtrapprox 5$~W in the entire range of wind speeds. The power dissipated by the probe and $R_3$ are the same, between $1$~W and $4.5$~W in the displayed airspeed interval.
}} \label{Pow}
\end{figure}

Lastly, there is an aspect which is not directly related to the measurement performance, but must be necessarily considered: it is the power consumption. Most of the electric power is dissipated by the arm of the bridge that conducts the probe current. In fact, from the total electrical current that circulates through the circuit, only a very small fraction is required for the controller operation. To obtain an estimate of the power required to keep the probe at a given temperature, 2D simulations using the finite elements method were performed. These allowed to know the rate of heat flow towards the surroundings at several airspeeds. In particular, at $u=16$~m/s the resulting power requirement was $3.6$~W, which is not too far from the measured value. From the calibration data displayed in the Figure \ref{KLP}, it can be deduced that at $u=16$~m/s the power dissipated by the probe is $P_\mathrm{p}\approx 4.5$~W. In the design reported here, the bridge and the controller power supplies are separated. This prevents the possibility of couplings between voltage fluctuations in the bridge supply and the controller supply. The path followed by most of the current delivered to the bridge is formed by the pass transistor, $Q_1$, and the resistors $R_3$, $R_c$, and $R_4$. Under normal operation, $R_4=R_3$. If we neglect the cable resistance, we see that the the probe arm of the bridge has a resistance given simply by $R_\mathrm{p}=2 R_3 \approx 4.6$~$\Omega$. The other arm has essentially a resistance $R_1+R_2=20$~k$\Omega$, so that the power it dissipates is four orders of magnitude smaller than that of the probe arm. Of course, the power dissipated by the anemometer will depend on the wind speed, and the voltage delivered by the power supply. Figure \ref{Pow} displays the power consumption of the bridge, the power dissipated by the pass transistor, $Q_1$, and the power dissipated by the probe, $R_4$, as functions of the wind speed $u$. The supply voltage is $V_\mathrm{CC}= 10$~V. The probe power is half that of the bridge, and the power dissipated by the transistor is relatively constant. We see that the values of the power dissipation are not small, so that adequate heat sinks must be used for the transistor $Q_1$ and the resistor $R_3$. Both must be rated to support levels of power greater than the maximum they must manage.
\\

\section{Conclusions}
\label{Concl}

We have designed and built a thermal anemometer appropriate for measurements of airspeeds from $0.2$~m/s through $16$~m/s within a frequency band from $0$~Hz through $100$~Hz. Its main advantage, as compared with hot-wire anemometers, is its robustness. This makes it especially adequate for outdoor measurements, in instances where the measuring of small scale turbulence is not required. It can also be used for closed loop control of the airspeed in wind tunnels. Of course, given that it works at low temperature, a simultaneous measurement of the air temperature is mandatory in order to correctly calculate the wind speed. Thus, a complete anemometric system based on this anemometer would comprise, in addition, a thermometer based on a calibrated thermistor or a thermocouple, and a two channel digitizer connected to a micro-controller or a PC, to perform the required real time computations. Of course, when the air temperature is known and constant, after a calibration it can be used without the need of temperature correction.

\begin{acknowledgments}
Financial support for this work was provided by FONDECYT Grant No. 1090686, and DICYT-USACH project No. 041231LM.
\end{acknowledgments}

\appendix

\section{On the utilization of alternative sensors}
\label{Ap_A}

A number probes based on the thermal conduction of heat to the surrounding fluid can be found in the literature. We will not consider here the hot wire, because the motivation of this work was precisely avoiding its use for monitoring and/or flow control purposes, given its fragility. Nevertheless, the interested reader will find an excellent account of its principles and applications in the book by Bruun.\cite{Bruun95} An excellent review of thermal anemometry, including hot film probes, is given by Fingerson.\cite{Fin94}

Virtually every electrical device capable of power dissipation can be used to implement a thermal anemometer, as long as one of its electrical parameters exhibits a noticeable---and strictly monotonous---dependence on the temperature. In this respect, the negative temperature coefficient (NTC) thermistor is possibly the device of choice: its resistance roughly changes as the exponential of the reciprocal of the temperature. This is probably the reason why relatively soon after its introduction in the marketplace, thermal anemometers based on thermistors were developed and studied.\cite{San50} When used to measure temperature, thermistors appear to be quite stable.\cite{Edwards83} In a thermal anemometer, they may have to withstand substantial amounts of current and power, possibly at an elevated temperature. This could degrade the characteristics of the device in a relatively short time, depending on the operating conditions. Also, due to the dependence of the resistance on the absolute temperature $T$, we have $\Delta R / R \sim -\Delta T / T^2$, so that at higher temperatures the signal to noise ratio worsens. For a thermal anemometer, the basic idea is to apply an electrical current so that the thermistor is self heated by Joule dissipation. Next, the operating mode will depend on the parameter that is held constant. At constant current, the fluid flow induces changes in the voltage drop between the terminals. The changes in this voltage inversely correlate with the changes in the flow speed. Conversely, at constant voltage the changes in the current inversely correlate with the changes in the flow speed. These are the two ``passive'' operating modes. A third realizable possibility is keeping constant the temperature. This, of course, requires a servo controller. The advantage of the latter mode is that---ideally---the thermal inertia of the probe plays no role. Thus, with an optimal servo controller, the response time is not penalized by heating or cooling times, contrary to what happen in the two former operating modes. Of course, these three principles can be applied to every thermal anemometer, not only to those based on the NTC thermistor. Now, in this particular probe the sensing element is normally encapsulated in glass, adding a thermal resistance in series to the thermal resistance between the surface of the enclosure an the surrounding fluid. This degrades the response time of the sensor: it introduces a time constant in exact analogy to adding an RC low-pass filter at the input of an amplifier. For monitoring of fluctuating flow speeds, this can go from innocuous through disastrous, depending on the value of the time constant. Commercial thermistors come in several formats. The fastest ones have no enclosure, and the bead size can be about $100 \mu$m. Their time constant in still air can be about $0.1$~s. Their terminals have a diameter of about $17\mu$m, so that they are fragile. In addition, they cannot dissipate large amounts of power, and their working temperature cannot be very high. They have been used for low speed measurements in liquids. Larger thermistors are robust, but their thermal time constant in still air can be as large as $20$~s. Of course, when the surrounding air is in motion, the response can be much faster. For example, a microthermistor with a $100\mu$m bead can respond to turbulent temperature fluctuations above $5$~kHz when the mean speed of the airflow is about $25$~m/s. These facts have implication for the operation of thermal anemometers. The bandwidth of any thermal anemometer depends on the mean speed of the fluid. In fact, this problem is even worse, because the dependence in the flow speed implies that the probe has nonsymmetric heating and cooling rates. At low flow speed, the cooling rate is also low, whereas the heating rate is determined essentially by the power source. In pulsating flows this can be a serious problem. Of course, an enclosure with high thermal resistance makes this problem even worse. Nevertheless, within reasonable limits, thermal anemometers based on thermistor can be designed to meet specific requirements. Examples of thermistor-based anemometers can be found in the works by Murphy and Sparks,\cite{MurSpa68} and Martino and McNall Jr.\cite{MarMcNall71}

Resistive temperature detectors (RTDs) have been extensively used to implement thermal anemometers. Perhaps their main advantage is a nearly linear dependence of the resistance on the temperature. Their drawback is that they are less sensitive to temperature changes. Nevertheless, they can be used at higher temperatures. This allows measurements less sensitive to the fluid temperature, provided that the difference between the working temperature of the CTA probe and the fluid temperature is large enough. For example, the working temperature of a hot wire CTA probe is typically about $300^\circ$C, which gives a difference of some $280^\circ$C when the fluid is air at room temperature. The platinum or tungsten wires used in HWAs fall in the RTD category, as well as the probe made of enameled copper wire reported here. Commercial, platinum based RDTs designed to be used as thermometers, have been used in thermal anemometers. The well known PT100 probe has been used in thermal anemometers and flow meters by several groups. See for example the work by Miete and Ray \cite{MieRay01}, where the performances of PT100, and p-n junction based flow meters are assessed. The PT100 sensor comes in several formats, like bars or chips, allowing their use in a variety of applications. It is characterized by a resistance $R = 100 \Omega$ at $T_{\mathrm C} = 0.00^\circ$C. At $T_{\mathrm C} = 90.0^\circ$C, its resistance is $R = 134.71 \Omega$. Around $T_{\mathrm C} = 90.0^\circ$C, we have that the change per K in the platinum resistance is $0.28$\%, whereas in the probe reported here it is $0.35$\%. Thus, the copper probe is $25$\% more sensitive than the PT100 probe, which should contribute to a slight improvement in the signal to noise ratio. As shown in Section \ref{Test}, the probe reported here seems to respond well up to $\sim 100$~Hz, according to the spectrum in Figure \ref{Spectrum}. This is due in part to the low thermal resistance between the surrounding air and the heating element. In fact, the enameled copper wire is physically in contact with the copper enclosure, even so, high temperature silicone was added to further increase the thermal conductance. Also, as mentioned in Section \ref{M_E}, the copper enclosure homogenizes the temperature along the probe. These two characteristics should contribute to a faster response of the copper probe. The ceramic enclosure has a lower thermal conductance than the copper enclosure, which leads to two degrading effects: i) the thermal coupling between the surrounding air and the heating element is weaker; and ii) axial thermal gradients could exist obeying to the same reason, degrading even more the probe performance. For probes of each type having the same geometry, the response time will be determined in the first place by the thermal mass of each probe, and the internal thermal resistance can add an additional lag, depending on how large its value is. A quantitative assessment of the possible advantages or drawbacks of each one of these probes is beyond the scope of this work. Clearly, an extensive program of testing should be necessary to establish the application fields where each probe could perform better.


%

\end{document}